\documentclass[useAMS,usenatbib]{mn2e}

\def\simlt{\lower.5ex\hbox{$\; \buildrel < \over \sim \;$}}
\def\simgt{\lower.5ex\hbox{$\; \buildrel > \over \sim \;$}}
\def\etal{{\it et al.}}
\def\etc{{\it etc}}
\def\ie{{\it i.e.}}
\def\eg{{\it e.g.}}
\def\cf{{\it c.f.}}

\def\pap3{Polar Shapelets}

\usepackage{graphicx}  

\title[CTI correction for HST/ACS]{Pixel-based correction for Charge Transfer Inefficiency in the Hubble Space Telescope
Advanced Camera for Surveys}
\author[Richard Massey \etal]
{Richard Massey$^{1}$\thanks{E-mail: rm@roe.ac.uk},
Chris Stoughton$^{2}$,
Alexie Leauthaud$^{3}$,
Jason Rhodes$^{4,5}$,\newauthor
Anton Koekemoer$^{6}$,
Richard Ellis$^{5}$
and
Edgar Shaghoulian$^{7}$\\  
$^{1}$Royal Observatory, Blackford Hill, Edinburgh EH9 3HJ, U.K.\\
$^{2}$Fermi National Accelerator Laboratory, P.O. Box 500, Batavia, IL 60510, U.S.A.\\
$^{3}$Physics Division, Lawrence Berkeley National Laboratory, Berkeley, CA 94720, U.S.A.\\
$^{4}$Jet Propulsion Laboratory, California Institute of Technology, Pasadena, CA 91109, U.S.A.\\
$^{5}$California Institute of Technology, 1200 East California Boulevard, Pasadena, CA 91125, U.S.A.\\
$^{6}$Space Telescope Science Institute, 3700 San Martin Drive, Baltimore, MD 21218, U.S.A.\\
$^{7}$Stanford University, Physics building, 382 Via Pueblo Mall, Stanford, CA 94305, U.S.A.}

\date{Accepted 2009 September 1.  Received 2009 August 31; in original form 2009 August 5.}

\begin{document}

\pagerange{\pageref{firstpage}--\pageref{lastpage}} \pubyear{2002}

\maketitle

\label{firstpage}

\begin{abstract}
Charge Transfer Inefficiency (CTI) due to radiation damage above the Earth's atmosphere creates spurious
trailing in Hubble Space Telescope (HST) images. Radiation damage also creates unrelated warm pixels --
but these happen to be perfect for measuring CTI. We model CTI in the Advanced Camera for Surveys (ACS)/Wide
Field Channel (WFC) and construct a physically motivated correction scheme.
This operates on raw data, rather than secondary science products, by returning individual electrons to 
pixels from which they were unintentionally dragged during readout.
We apply our correction to images from the HST COSMOS survey, successfully
reducing the CTI trails by a factor of $\sim30$ everywhere in the CCD and at all flux levels. We 
quantify changes in galaxy photometry, astrometry and shape. The remarkable $97\%$
level of correction is more than sufficient to enable a (forthcoming) reanalysis of
downstream science products, and the collection of larger surveys.
\end{abstract}


\section{INTRODUCTION}

Charge-Coupled Device (CCD) imaging detectors convert incident photons into electrons. The electrons are stored within a
silicon substrate, in a pixellated grid of electrostatic potential wells that gradually fill up during an exposure. At the end
of the exposure, the electrons are shuffled, row by row, to a readout register at the edge of the device. They are then counted
and converted into a digital signal.

\begin{figure*}
\includegraphics[width=160mm]{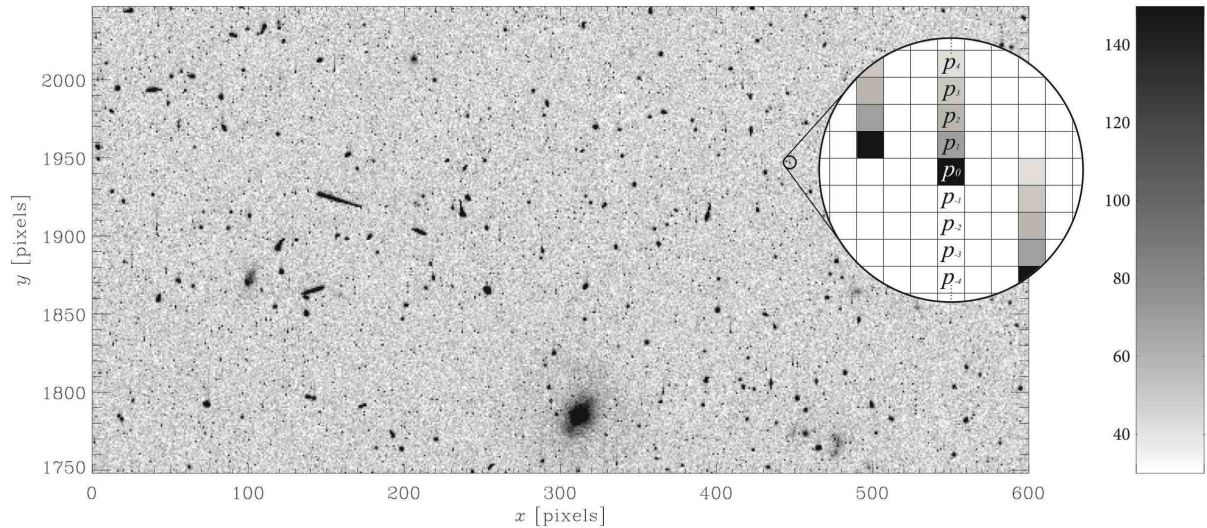}
\caption{A typical, raw HST ACS/WFC image, in units of electrons. 
This 30$\arcsec\times$15$\arcsec$ ($600\times 300$ pixels) 
region is at the far side of the CCD to the readout register,
which lies towards the bottom of the page. It was
obtained on 15 May 2005, 1171 days after the launch of ACS. Upon
close inspection, as illustrated in the zoomed inset, 
the CTI trailing behind (above) objects is manifest.}
\label{fig:rawacsimage}
\end{figure*}

Above the Earth's atmosphere, a continuous bombardment of high energy particles
makes a harsh environment for sensitive electronic equipment. Protons, neutrons,
high-energy electrons and heavy ions create bulk damage by colliding with and
displacing atoms from the silicon lattice. 
The dislodged atoms can come to rest in the interstitial space, and the
vacancies left behind move about the lattice until they combine with interstitial impurities,
such as phosphorous, oxygen or another vacancy
\citep{janesick01}. 
Such defects degrade a CCD's ability to shuffle electrons,
known as its Charge Transfer Efficiency (CTE; Charge Transfer {\em In}efficiency
CTI=1-CTE). Electrons can become temporarily trapped in the local potential, then released
after a delay that depends upon the
properties of the lattice and impurities, and the operating temperature of the 
device \citep{srhsr,srhh}. If a few electrons are
trapped during CCD readout, and held while other electrons are moved along, they are released as a
spurious trail (see figure~\ref{fig:rawacsimage}). Regions of the image 
furthest from the readout register are worst affected, because electrons starting
there encounter the most charge traps during their journey across the device.

Charge Transfer Inefficiency gets worse over time, as cumulative radiation
damage creates more charge traps. By January 2007, when the Hubble Space
Telescope (HST) Advanced Camera for Surveys (ACS) encountered an electronic
failure, the instrument had been in operation on-orbit for almost 5 years and
its Scientific Imagaging Technologies (SITe) ST002A CCDs were already significantly affected by CTI. Similarly, the Wide Field
and Planetary Camera 2 (WFPC2) instrument spent 16 years on-orbit. By the time it
was decommissioned, its truly severe CTI created trailing that was readily
visible on every exposure.

Different astronomical measurements are hindered by different species of charge traps.
Charge traps with a characteristic release time of a few CCD
clock cycles move electrons by a few pixels, creating short trails that alter
the apparent position (astrometry) and shape (morphology) of faint objects.
Charge traps with a release time of many clock cycles completely detach
electrons from their original objects, thus also lowering their measured
brightness (photometry).
Several parametric schemes have been developed to estimate and thus correct all these effects 
at a catalogue level. Fitting formulae have been approximated, as a function of objects'
detector position, date of observation and flux, for incorrect astrometry with WFPC2
\citep{cawley02}, photometry with ACS \citep{riess03}, morphology with STIS
\citep{rhodes04} and ACS \citep{rhodes07}, and others. Although such schemes provide
first order mean correction for a large number of objects, they are 
inaccurate for high precision measurement of individual objects. 
For example, no account is made for the screening of faint objects by
bright objects slightly closer to the readout register (which pre-fill charge traps) or,
with extended sources, for the dependence upon object size, radial profile, and shape
(see discussion in \citealt{riess00} and \citealt{cti2}).

More precise removal of CTI trailing requires manipulation of the raw pixel data, using software to move
electrons back to their original locations. Since CCD readout is the last stage of data acquisition, this
should be the first stage of a final data reduction pipeline \citep[\eg\
CALACS,][]{acshandbook}. Even this will never achieve precision at the level of single electrons,  since
charge trapping and release are stochastic, quantum mechanical effects. However, such software has been
demonstrated to be much more accurate than parametric solutions for the correction of HST STIS imaging
\citep[][2006 \& 2007]{bristow03im,piatek05} and spectroscopy \citep{bristow03sp}, and Chandra ACIS spectroscopy \citep{grant04}.
This approach can also provide physically motivated model parameters, rather than an ad-hoc fitting
function to arbitrary parameters.

In this paper, we develop an empirical but physically motivated, pixel-based CTI correction scheme for ACS/WFC
imaging. We concentrate on traps with release times of a few CCD clock cycles, but our algorithm is
sufficiently general to be able to incorporate additional species of traps with longer release times if
necessary. In \S\ref{sec:model}, we build a model of the
readout process by measuring the rate at which the CCDs' potential wells are filled by
electrons, and the properties of charge traps within those wells. 
In \S\ref{sec:correction}, we describe software to implement this readout model, and use
the iterative approach of \citet{bristow03im} to reverse the process. In \S\ref{sec:tests}, we
apply this to real data and obtain better measurements of galaxy photometry, astrometry and ellipticity in the
HST COSMOS survey. In \S\ref{sec:conclusions}, we discuss possible improvements to our algorithm, and summarise our work.

\section{MODELLING THE ACS/WFC CCDs} \label{sec:model}

\subsection{Data and dates} \label{sec:data}

We primarily base our analysis on the ``{\tt \_raw}'' data obtained during HST
cycles 12 and 13 for the COSMOS survey (HST-GO-09822, P.I.: N.\ Scoville). These
are 2368 uniform, extragalactic exposures of 507~seconds each
\citep{scoville07,koekemoer07}. For convenience,
we first split each exposure into the four quadrants read out through different
amplifiers, rotating and flipping them to orient the readout in the same
direction for all. We then multiply the image by the calibrated gain 
(which varies slightly between amplifiers) and subtract
the corresponding superbias images created by STScI. This reverts the images to
units of electrons, in approximately their configuration on the CCD immediately prior to readout
(but note uncertainty about the imperfectly 
modeled injection of charge by warm pixels during readout in \S\ref{sec:warmpixels}).

We have also tested our CTE correction method in a less systematic manner on
imaging from cycle 14 (HST-GO-10496, P.I.: S.\ Perlmutter). This extends the
analysis to nearly January 2007, when ACS failed. While this smaller survey
provided sufficient data to verify an extrapolation of our correction
parameters, such non-uniform observational strategies proved less useful for
tracking the gradual CTE degradation.

Our model could also be extrapolated to correct observations obtained after
servicing mission 4. However, this would be a large extrapolation, and the trap density did
not necessarily continue to rise at the same rate while the instrument was offline.
We would therefore recommend new calibrations as soon as sufficient images become available.

\subsection{Warm pixels} \label{sec:warmpixels}

A detector's CTI can be measured using First Pixel Response (FPR) tests, Extended Pixel Edge Response (EPER) tests,
or $^{55}$Fe radiation \citep{janesick01}. FPR and EPER tests involve uniformly illuminating a CCD and measuring
deviations from that uniformity in the few pixels nearest the readout register and (virtual) pixels obtained by
continuing to  clock charge past the end of the CCD. In addition, a radioactive iron source can be used to create a
series of $\delta$-functions on the CCD, all of a fixed energy level. Deficiencies in the observed number of
electrons in some events, or deviations in their shape, reveal the presence of charge traps between the event and the
readout electronics. 

A continuous programme of FPR and EPER monitoring has been carried out on the ACS/WFC detectors since launch
\citep[HST programs 9649,10044,10369 and 10732;][]{mutchler05}. 
Radiation testing was performed prior to launch, but is not possible on-orbit in HST.

After experimenting on all three types of data acquired in a laboratory environment, 
we decided that that the parameters needed for our model are most
easily extracted from radiation tests. Although in-flight $^{55}$Fe data are unavailable, cosmic ray
damage has also created randomly distributed ``warm pixels'' that suffice instead \citep{biretta05}. Warm
pixels are short circuits in the CCD electrostatic potentials used to collect charge, and continuously
inject spurious charge into the device. During a long exposure, warm pixels create single-pixel
$\delta$-functions. Since the rate of charge injection varies between warm pixels, these $\delta$-functions
have a range of amplitudes. In some ways, this is actually more convenient than an $^{55}$Fe source.

To identify warm pixels within each image $I$, we locate 2D local maxima, after unsharp masking to
eliminate extended objects (whether they be resolved galaxies or merely a Point Spread Function). One
such warm pixel is labeled $p_0$ in figure~\ref{fig:rawacsimage}, and contains $n_e=I(p_0)$ electrons. This
number has been only slightly changed during readout. We measure the trail
\begin{equation} \label{eqn:trail}
T_i(n_e)\equiv I(p_{i})-I(p_{-i})~,
\end{equation}
\noindent where $p_{\pm i}$ are pixels defined relative to the location of the
warm pixel $p_0$ in figure~\ref{fig:rawacsimage}, and $i$ is an integer from $i=1...9$. Note that 
we could have replaced pixels with negative indices by a locally-determined background level. Our better scheme
is robust to the accidental inclusion of any extended sources 
(or cosmic rays covering more than one pixel) in our catalogue. 
Although such individual artefacts have a shape, they should not have a preferred
direction on average, and therefore should not bias our measurement of $T_i$. Indeed, we could even have
intentionally measured the trailing behind extended sources. However, in practice, averaging out their
intrinsic radial profile adds more noise than the extra numbers of pixels add signal. 

At the very faint end of our warm pixel catalogue, which we push down to $n_e=100$~electrons so that it only just exceeds the
background level, our algorithm often finds noise peaks instead of true warm pixels. If the noise were due to photon shot noise,
these pixels would be ideal for our purposes, but it is more often read noise -- which is added after clocking and therefore
untrailed.  To prevent the dilution of our trail measurements, we discard pixels that are not flagged as warm in at least half of the
exposures. In the faintest two bins,  this process removes $85\%$ and $50\%$ of the catalogue, but quickly becomes negligible for
brighter pixels. True warm pixels persist through many exposures.

We avoid ``hot'' pixels containing more than 76230~electrons, which is $\sim90\%$ of the full well depth.
Saturated pixels bleed charge into nearby pixels on the same column,
interfering with the observed shape of the CTI trail. 

Note that warm pixels also inject a low level of charge into a column of pixels during readout. This is a
relatively small complication: the ratio of charge in this line to that in the $\delta$-function is equal
to the ratio of the clock speed to the total exposure time. The data can be corrected along all but the
worst-affected columns by subtracting superbias frames (produced from a zero-second exposure but normal
readout). We subtract bias frames before anything else, since we are trying to obtain data as it was
immediately before readout. However, this is a simplification because (some of) the spurious charge
was also present during readout, when the CTI trails were created. A more complete CCD model might also
measure the ``temperature'' of each warm pixel, then incorporate a continuous injection of charge during
readout. This would be most important for measurements of the density of charge trap species with
long release times, whose extended trail is partly degenerate with a constant line of charge injection
(as opposed to charge re-emitted from short timescale traps for which the original source can be readily identified).
Even more worryingly, there is (unexplained) temporal structure in the injection of charge to some warm
pixels that is not necessarily reproduced in the superbias frames.
We therefore decided to incorporate the few percent of columns containing the warmest pixels in the
CALACS bad pixel mask. This is more severe than excluding hot pixels, since the entire rest of the column is masked.

\subsection{Model assumptions and parameters} \label{sec:assumptions}

To emulate the process of CCD readout, we first need to model the collection of electrons within a pixel's
potential well. In theory, the expected locations of a given number of electrons can be calculated from a model
of the CCD, by solving Poisson's equation within the appropriate applied potential. \citet{hardy98} and
\citet{seabroke} find that, for typical CCDs, electrons are contained within a well-defined volume -- outside
which their density falls rapidly to zero. Adopting this result, we shall regard a given number of electrons
as filling a three dimensional pixel up to a specific height.

We then need to model the charge traps. We assume that none are filled during integration. However, 
following {\eg} \citet{grant04} \citep[but not][]{bristow03im}, we
simplify the problem of charge capture as soon as readout begins by assuming that unoccupied traps 
{\it instantly} capture any electrons
in their vicinity (\ie\ all traps inside the well-defined volume get filled). In reality, there is an
exponential decay time needed for traps to capture an electron, but this is typically much shorter than the
electron dwell time within one CCD clock cycle \citep{hardy98}. The rapid and efficient capture of electrons, even
in locations where the electron density is low, further delineates the quantifiable volume of a pixel occupied
by a given number of electrons.  It also means that any charge traps physically located near the bottom of a
pixel (below the ambient background level) reach the steady state of being
permanently occupied, even during readout. In our model such traps never affect an image, so we can never probe them; but
then, we never need to.

Each filled trap releases its electron after a time determined by $\tau$, the e-folding time of an exponential
decay \citep{srhsr,srhh}. If the trap is inside the volume currently occupied by electrons, it will immediately
capture a new one; if it is above, the electron will be pulled down and added back to the cloud (which thus
expands and may even be exposed to new charge traps). We concentrate on species of traps with capture times
$\tau$ of a few CCD clock cycles ($3212\mu$s in the parallel direction; \citealt{sirianni04}). Traps with long
release times affect the leading edge of the first bright object in a CCD column, but then reach a full steady
state (at least while the rest of the object passes by) and only minimally affect its photometry, astrometry and
morphology \citep{cti2}. For a fixed clock speed, $\tau$ can be conveniently expressed in units of the number of
pixels that have passed since capture.

If the charge traps are evenly distributed, we need measure only the
mean density of each species\footnote{This may not be acceptable for radiation-hardened
detectors in future spacecraft that suffer very few traps \citep[see][]{dawson08}. In that regime,
techniques like pocket pumping would be ideally suited to measuring the  locations of
individual traps.}. In \S\ref{sec:volume}, we indeed find the distribution of traps to be  uniform in the
$y$ direction; we have no reason to suggest that the distribution is not similarly uniform in
the other directions. In addition, our approach will internally parametrize out any smooth
density gradients in the vertical direction. 

We assume that each charge trap always captures a single electron. Real traps might be able to hold several, or occasionally capture none. If they hold several, and especially if the trap densities are low, simultaneous release of many electrons could produce correlated spikes in the trails. Trapping mechanisms are not yet sufficiently understood to know whether electrons are released
singly or in groups (C.\ Bebek, priv.\ comm.). We therefore simply assume that any multiple (or partial) occupancy of
traps is equivalently modelled by an the {\it effective} density of single-electron traps. 

Finally, we ignore CCDs' three-phase clocking cycle, by which a cloud of electrons is transferred from one
pixel to the next \citep{janesick01}. The transition period exposes the electron cloud to additional
regions of silicon, and additional charge traps. However, with our assumption of instantaneous charge
capture, extra charge traps in regions between pixels can again be modeled to first order as an increase in the
{\it effective} trap density. These traps are a fraction of a pixel nearer to (or further from) the readout register,
but horizontal translations of the exponential release curve are also degenerate with a change in normalisation.
The only remaining finesse is that the width (and therefore the height) of the electron
cloud changes during the three-phase clocking cycle, making the cloud dwell temporarily at different
heights within the silicon and therefore being exposed to a different number of charge traps.
We ignore this second-order effect, but incorporating a model of the full, three-phase clock cycle might be a useful exercise for future work.

Our model features the following free parameters:
\begin{itemize}
\item number of (relevant) species of charge trap,
\item characteristic release time of each species of charge trap,
\item density of each species of charge trap
\end{itemize}
(or a functional form for their distribution if they are not uniformly located throughout the CCD), 
plus the
\begin{itemize}
\item height occupied by a cloud of electrons in a pixel's potential well
(\ie\ the number of charge traps they are exposed to), as a function of the number of electrons.
\end{itemize}
As we shall now demonstrate, warm pixels enable us to measure all of these quantities from on-orbit science exposures.

\begin{figure}
\includegraphics[width=84mm]{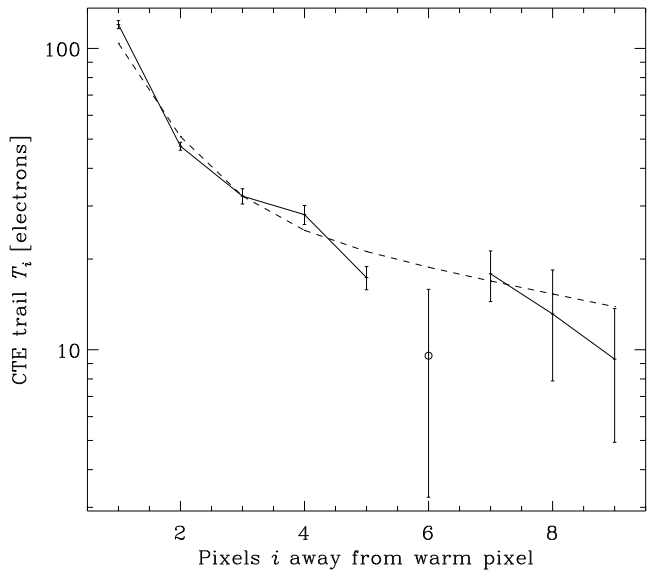}
\caption{The mean CTI trail behind 63901 warm pixels containing between 3234 and 76230 electrons each, from 20 exposures obtained 1171
days after the launch of ACS. Error bars show the $1\sigma$ scatter from measurements behind the whole population of warm pixels. 
The dashed curve shows the best-fit double exponential model.}
\label{fig:onetrail}
\end{figure}

\subsection{The characteristic release times of different species of charge trap} \label{sec:species}

To obtain maximum signal to noise, we begin by studying the brightest warm pixels, furthest from the readout register and in data
obtained near the chronological  end of the COSMOS survey. We select 20 images taken on 15 May 2005, 1171 days after the launch of ACS,
and measure the mean trail $T_i$ around all warm pixels between 1634 and 2039 pixels from the readout register, and containing between
3234 and 76230 electrons\footnote{These values are chosen so that more bins can be added later, equally spaced in distance from the readout
register and log($n_e$).}. The mean trail is shown in figure~\ref{fig:onetrail}.

\begin{figure*}
\includegraphics[angle=90,width=128mm]{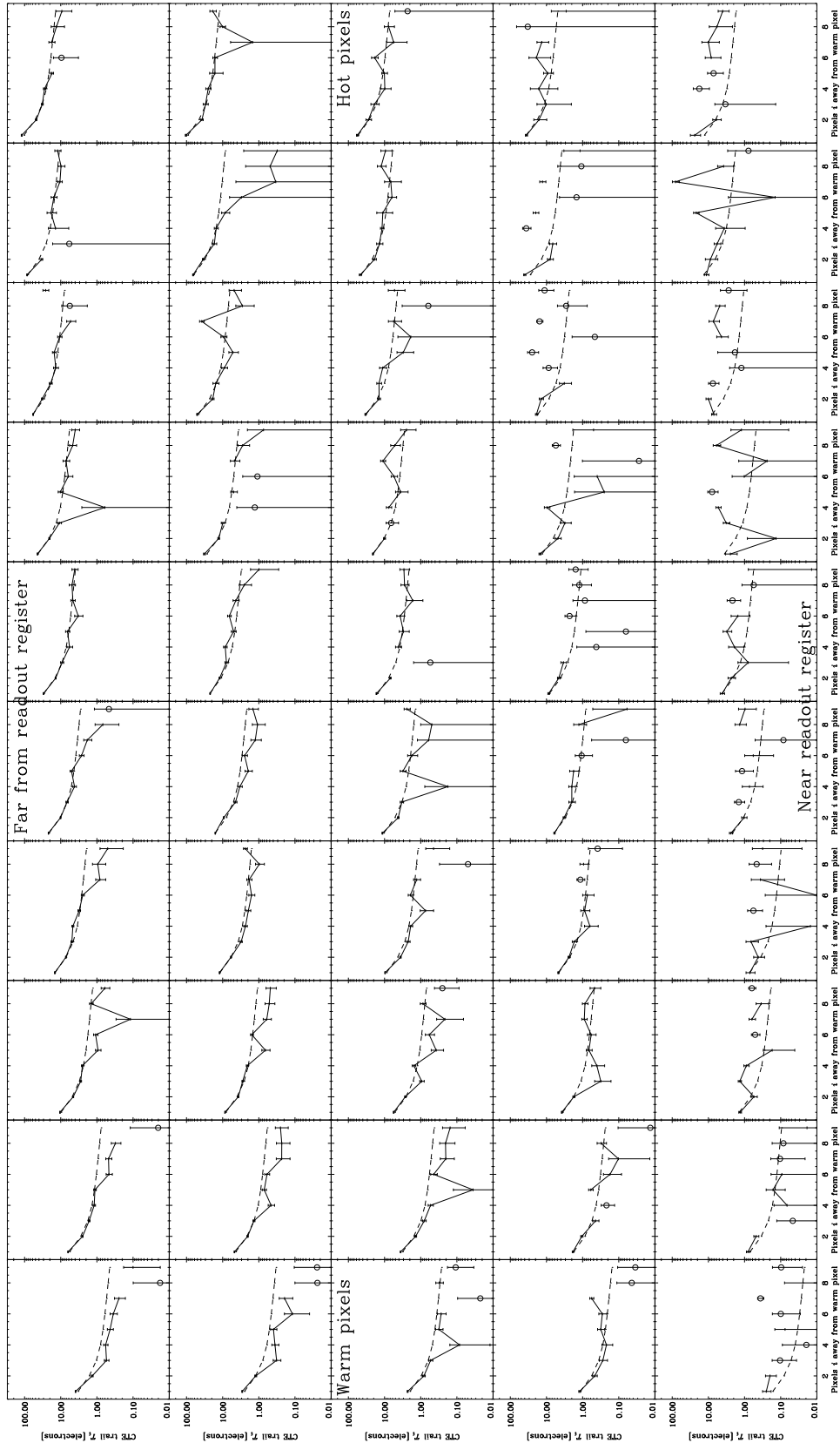}
\caption{CTI trails behind a range of warm pixels, containing different amounts of flux
(increasing left to right) and at greater distances from the readout register (increasing bottom
to top). Figure~\ref{fig:onetrail} is reproduced exactly in the top right panel.
The smooth dotted lines show fits of double exponential function~\ref{eq:doubleexp}, assuming a fixed
ratio of the densities of different species of charge trap, and hence the same shaped curve in each case.
Open circles depict negative values.
These measurements can be used to probe the distribution of charge traps on the CCD, and the
rate at which a pixel's potential well fills up as a function of the incident flux, thereby
exposing electrons to additional traps. The mean distances from the readout register for each row,
and the flux levels for each column, are provided in figure~\ref{fig:qn}.}
\label{fig:rawtrails}
\end{figure*}

Using Levenberg-Marquardt least-squares minimization, we find that the mean trail is well-fit by a
sum of two exponentials,
\begin{equation}
T_i=A_1e^{-i/\tau_1}+A_2e^{-i/\tau_2}~,
\label{eq:doubleexp}
\end{equation}
\noindent where $A_1=327\pm6$~e$^-$, $\tau_1=10.4\pm3.2$~pixels, $A_2=108\pm3$~e$^-$ and
$\tau_2=0.88\pm0.2$~pixels. 
At the $-83^\circ$C operating temperature of ACS/WFC, these correspond precisely with the release times of trap species 
found by \cite{hopkinson01} in a Marconi CCD47-20 n-channel detector at 0.34~eV and 0.31~eV below the conduction band.
The origin of these particular traps is unknown, but they have been speculated to be E-centres (Phosphorous-Vacancy complexes) with
impurities of either carbon \citep{hopkinson91} or hydrogen \citep{tokuda00}.
Pure E-centre charge traps at 0.44~eV exist in both the Marconi and ACS/WFC \citep{jones00} CCDs, with a 
characteristic release time of $\tau_\mathrm{E}\sim300$~pixels in the Marconi. 
All other charge traps in the Marconi CCD have $\tau<0.01$~pixels \citep{hopkinson01}, so this 1-to-1 correspondence is
unlikely to be coincidence.

However, allowing a third exponential prevents our fit from converging, with positive $A_3$ but $\tau_3$
iterating towards infinity. Fitting slowly decaying exponentials is a standard problem \citep[\eg][]{cover08}, because they
are nearly degenerate with a constant. A line of charge above a warm pixel could represent either a species of trap with
a very long release time or contamination from continued charge injection during readout
(\cf~\S\ref{sec:assumptions}). We therefore restrict our analysis 
to the two species of traps with shorter $\tau$, which also have a much more profound effect on object astrometry and morphology \citep{cti2}. 
Henceforth, we shall fix
the release times of the charge trap species at $\tau_1\equiv10.4$ and $\tau_2\equiv0.88$~pixels, and also fix 
their relative densities to be 3.0:1\footnote{There is no 
reason why we would {\it expect} this to be an integer.}.
Fitting to earlier ACS data yields consistent (but noisier) values of $\tau$ and the relative density.

\subsection{Height of pixel occupied by electrons and the density of charge traps}
\label{sec:volume}

The total number of electrons that have been displaced from a packet originally
containing $n_e$ electrons is the integral under the curve 
\begin{equation} \label{eq:ntraps}
\sum_1^\infty T_i=\frac{A_1}{e^{1/\tau_1}-1}+\frac{A_2}{e^{1/\tau_2}-1}~\bigg[=n_q\bigg]~.
\end{equation} 
\noindent Since we are assuming that any traps surrounded by free electrons are
instantly filled, this number of trapped electrons equals the (effective) number of charge traps $n_q$ in
the volume $V$ traversed by electrons between their original pixel and the
readout register. 

The volume $V$ is the product of the distance $y$ to the readout register, the width of a pixel, and the height within a
pixel\footnote{This assumes that the height of an electron cloud changes by only a small amount during readout; for
further discussion, see \cite{cti2}.} filled by $n_e$ electrons, above a background level $b$. Warm pixels are available
with a range of $y$ and $n_e$ and in images with a range of backgrounds $b$ (this varies around $51\pm9$~e$^-$ due to
the angle between the telescope and the sun: \citealt{leauthaud07}). By counting the number of electrons trailed behind
a variety of warm pixels, we  directly measure $n_q(y,n_e,b)$. Note that the interpretation of electrons filling pixels
to a given height is instructive but not necessary: this function is precisely what we require.

Figure~\ref{fig:rawtrails} shows the mean CTI trail behind warm pixels at varying distances from the readout register
and of varying warmth. The top-right panel reproduces figure~\ref{fig:onetrail}. The other panels show trails behind
different pixels within the same set of 20 images. Although a busy plot, it has been simplified for illustration. For
the real analysis, the number of bins was doubled in both $y$ and $n_e$ directions, and an additional dimension, with
five bins of varying background level, has been marginalised over here. 
Towards the left hand side, the trails may be systematically underestimated due to the inclusion in our catalogue of
read noise peaks. However, the efforts in \S\ref{sec:warmpixels} to exclude them have kept this small, as demonstrated
by the still-recovered shape of the trails. 
Each trail is fitted with double exponential decays (\ref{eq:doubleexp}), the integral under which gives the total
number of trapped electrons, and hence the number of exposed traps $n_q$.

\begin{figure}
\includegraphics[width=84mm]{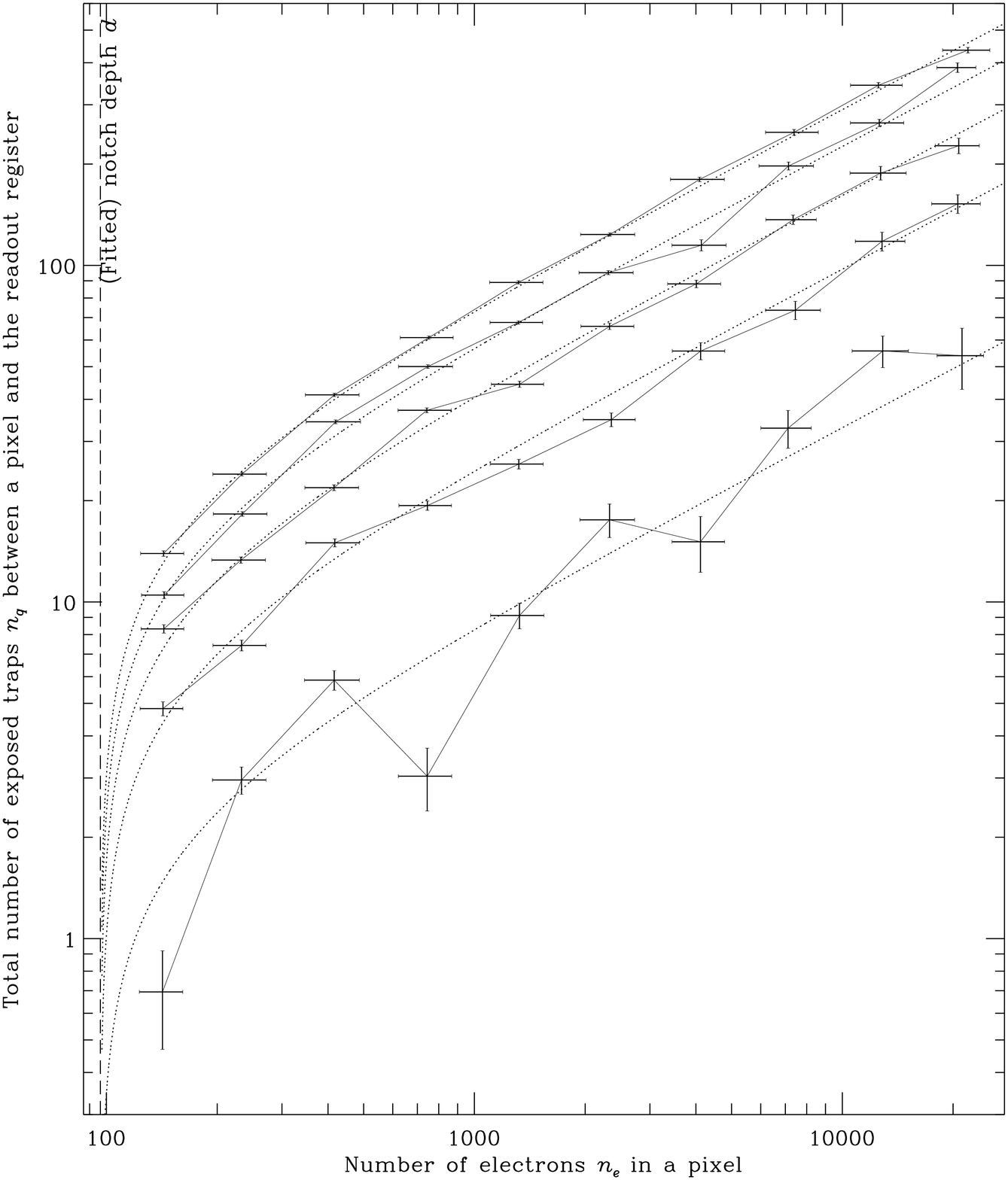}
\caption{The total number of charge traps $n_q$ exposed between a warm pixel and the readout register, as a
function of the number of electrons $n_e$ in that pixel. 
If traps are uniformly distributed, this is proportional to the height within a pixel filled by $n_e$ electrons.
Error bars are underestimated, because they do not
include contributions propagated from uncertainty in $\tau$. 
The solid lines join data from warm pixels 212, 618, 1024, 1430 and $1836\pm203$
pixels from the readout register (bottom to top). 
The cumulative number of traps increases linearly with that
distance, implying that the traps are uniformly distributed throughout the CCD.
The dashed curves show a model in which the electrons occupy zero volume within the CCD notch 
(the depth of which is a free parameter), and increase as a power law of the number of electrons above it.}
\label{fig:qn}
\end{figure}

Figure~\ref{fig:qn} shows the total number of charge traps exposed to each set of warm pixels for which a panel was
allocated in figure~\ref{fig:rawtrails}. Solid lines join the data points measured at constant distances from the 
readout register (rows in figure~\ref{fig:rawtrails}). 
These data are well modelled by
\begin{eqnarray}
n_q=\rho_qV~~~~~~~~~~~~~~~~~~~~~~~~~~~~~~~~~~~~~~~~~~~~~~~~~~~~~~~~~~~~~\nonumber\\
=\rho_q\left[\left(\frac{\mathrm{max}\{n_e-d,0\}}{w}\right)^\alpha
                -\left(\frac{\mathrm{max}\{b-d,0\}}{w}\right)^\alpha\right]~y^\beta~,
\label{eq:ctemetamodel}
\end{eqnarray}
where $\alpha$, $\beta$, $d$ and $\rho_q$ are to be measured. 
Parameter $w$ is the full well depth, which we assume to be fixed at 84700~e$^-$ even though 
it varies by a few thousand electrons across the device \citep{gilliland04}, and $d$
is the depth of the supplementary buried channel or ``notch'' in the CCDs, which we fit. The
notch is a small region at the bottom of a potential well used to gather
electrons when only a small number are present, much like a narrow channel cut
into the bed of a water drainage canal. In our model, the notch has
zero volume, so it merely adjusts the starting
point of the power-law increase in height governed by $\alpha$.

We again use Levenberg-Marquardt least-squares minimization to fit the free parameters.
We find that $\beta=1.01\pm0.01$, consistent with a
uniformly increasing number of traps with increasing distance from the readout register.
Since we would expect this behavior, we explicitly fix $\beta\equiv1$ and then obtain best-fit values of
$\alpha=0.576\pm0.013$, $d=96.5\pm2.0$~e$^-$ and $\rho_q=0.544\pm0.008$~pixel$^{-1}$, split between the two species of
charge trap in a ratio of 3.0:1. Note that these errors do not include the uncertainty propagated from the
measurement of the charge trap release times and density ratio.

\subsection{Growth rate of charge traps} \label{sec:growthrate}

ACS was installed on 7 March 2002, and cumulative radiation damage since then has gradually created more charge traps.
We can use the COSMOS data, which was acquired uniformly over a long period, to track CTI degradation. Assuming that
charge trap release times and the relative densities of the two species remain unchanged, we use the warmest pixels
furthest from the  readout register (as in \S\ref{sec:species}) to measure the density of charge traps $\rho_q$ at
twelve additional times during the survey.

\begin{figure}
\includegraphics[width=84mm]{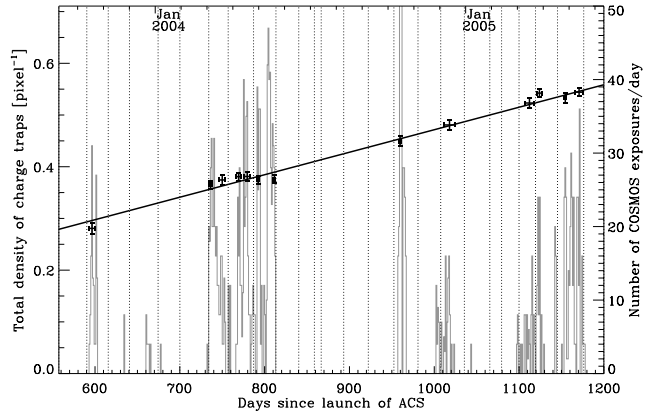}
\caption{The accumulation of charge traps in the ACS/WFC CCDs between 9 September 2003 and the end
of 2005. The data points and left hand axis show measurements of the total density of charge
traps, split between the two species in a ratio of approximately 3.0:1. Horizontal error bars show
the total time during which it was necessary to average data to obtain sufficient signal to noise.
The histogram and right hand axis show the number of exposures taken as part of the COSMOS survey,
which we analysed to obtain these figures. The dotted vertical lines show the times when the ACS
CCDs were annealed (these are also listed in table~\ref{tab:acsanneals}).}
\label{fig:trapgrowth}
\end{figure}

Figure~\ref{fig:trapgrowth} shows the measured increase in charge trap density over time. This is fit by a constant accumulation of
$(4.34\pm0.13)\times10^{-4}$ traps per pixel per day, and an initial density on launch of $\rho_q^0=0.037\pm 0.001$ traps per pixel 
(although the formal error on the latter is probably an underestimate, since the extrapolated value is very sensitive to errors in the
gradient). We also tried fitting a sawtooth model in
which a fraction of the traps were removed during their first subsequent anneal
(as described by \cite{cox01}, the temperature is raised to $13^\circ$C about once a month; 
the exact dates during cycles 12 and 13 are recorded in the appendix).
\cite{riess02} and \cite{sirianni04} confirmed that this successfully reduces the density of hot pixels.
However, the fraction of removed charge traps iterated to zero during our fit: recovering a linear model
\citep[consistent with][]{mutchler05} and suggesting that the anneals did not significantly improve CTE.
Indeed, while hot pixels and charge traps are likely both a product of the same non-ionising energy loss and should
accumulate at the same rate, \cite{robbins00} find that traps are only effectively removed by annealing at temperatures above
$150^\circ$C for E-centre traps and $330^\circ$C for divacancy traps.

\section{CCD READOUT SOFTWARE} \label{sec:correction}

\subsection{Forward operation} \label{sec:forward}

We now describe a software imitation
of the hardware CCD readout, using the model
developed in \S\ref{sec:model}. Just like the real version on HST, our software uses a fairly simple
algorithm to shuffle charge to an adjacent pixel, and this is repeated many times to complete a full
readout. For convenience, our code has a user interface written in IDL, but for speed, the core is
written in Java. Even so, the software algorithm is much slower than an on-board hardware readout,
taking 25~minutes per $4096\times 4096$ ACS/WFC image on a single 2~GHz processor, compared to
2.5~minutes on HST  \citep{acshandbook}. However, it is trivially parallelisable for multiple
exposures, and need ever be run only once on any given exposure.

Since the mean density of charge traps $\rho_q$ is known, but their exact locations are not, 
traps are scattered at random 3D locations within the model silicon wafer. Importantly, each
trap is allocated a particular vertical height in some detector pixel. The occupancy of each will be
continuously monitored. Electrons are then added into the pixel array, in the numbers we think they
were at the end of an exposure. 

A cloud of $n_e$ electrons is assumed to fill a pixel to a fractional height
$\left({\rm max}\{n_e-d,0\}/w\right)^\alpha$. 
Any charge traps inside the occupied volume immediately capture one electron.
All free electrons are then transferred instantaneously to their adjacent pixel (or to the serial register
for the last pixel in parallel array, or to the preamplifier from the last pixel of the serial
register). As discussed in \S\ref{sec:model}, we do not attempt to model the three-phase clock
cycle of a real CCD. Finally, electrons inside traps are allowed to decay with probability
$1-e^{-1/\tau}$. Released electrons are returned to the free electron pool, and the process is
repeated. 

Shot noise in this model created ragged trails behind faint
objects and amplified background noise fluctuations. We introduce two numerical schemes to ameliorate this.
Firstly, to counter the random release or non-release of electrons from charge traps, our model traps are allowed to contain
fractional numbers of electrons. In this scheme, each electron is gradually released in a floating-point exponential trail,
until the trap is refilled to a whole electron by the passing of a new bright pixel (removing only a fractional number of
electrons from the free pool). After readout, the number of electrons in each pixel is rounded back to an integer value for
storage.
Secondly, to counter the necessarily random locations of model traps, we can run the code multiple times with different random
seeds and average the results. To save CPU time, this is equivalently implemented by increasing the density of traps by a
factor of 3 (or any other number specified as an optional input parameter), each of which is only allowed to contain a third of
an electron. This most improves the sampling of traps in the thin slice of the CCD above the image background level that
profoundly affect faint sources. 
With both schemes, the change in background noise level during readout becomes less than 1\%.

\subsection{Reverse operation} \label{sec:reverse}

Moving electrons back to where they belong, and thus removing the CTI trails,
requires the image mapping to be inverted. This is not analytically possible.
However, the trailing represents a small perturbation around the true image, so
it can be achieved using the forward algorithm, via an iterative approach.
Following \cite{bristow02}, \cite{bristow03im} and \cite{bristow05}, 
table~\ref{tab:bristowmethod} describes a way to obtain an image that, after a
software readout, reproduces the data actually downloaded from HST. This is the
desired, corrected image. Images (A), (C), (E), \etc\ can all be obtained and,
after several iterations, converge to that ideal.

\begin{table}
\begin{center}
\begin{tabular}{llc}
\hline
\hline
True image              & \multicolumn{2}{l}{$I$~~~~~~~~~Not available}   \\
Downloaded from HST     & $I +\delta$                  & (A) \\
After one extra readout & $I+2\delta+\delta^2$         & (B) \\
(A)+(A)$-$(B)           & $I        -\delta^2$         & (C) \\
\hline
After another readout   & $I+\delta-\delta^2-\delta^3$ & (D) \\
(A)+(C)$-$(D)           & $I+\delta^3$                 & (E) \\
\hline
After another readout   & $I+\delta+\delta^3+\delta^4$ & (F) \\
(A)+(E)$-$(F)           & $I-\delta^4$                 & \etc \\
\hline
\hline
\end{tabular}
\caption{Iterative method to remove CTI trailing, using only a forward algorithm that \emph{adds} trailing.
The true image $I$ is desired but not available.  Only a version with (a small amount of) trailing, $I+\delta$, can be
obtained from the telescope.  However, by running that image through a software version of the readout process, and
subtracting the difference, deviations from the true image can be reduced to $\mathcal{O}(\delta^2)$. Successive
iterations further reduce the trails until an image is produced that, when ``read out'', reproduces data arbitrarily
close to those obtained from the telescope.  This is the corrected image.}
\label{tab:bristowmethod}
\end{center}
\end{table}

\begin{figure*}
\includegraphics[width=160mm]{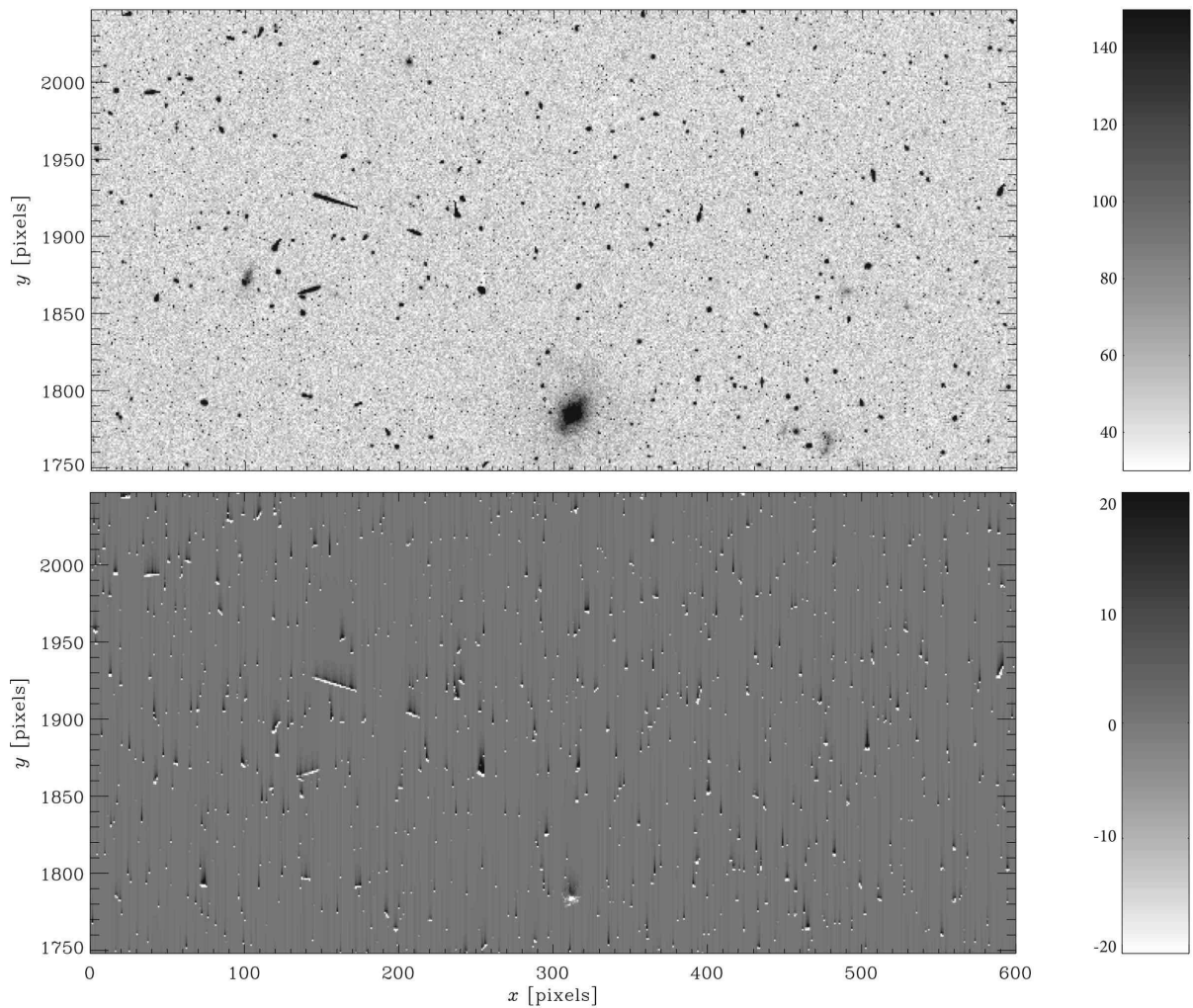}
\caption{{\it Top}: The HST ACS/WFC image from figure~\ref{fig:rawacsimage}, 
after CTI correction, in units of electrons. {\it Bottom}: Difference image.}
\label{fig:correctedacsimage}
\end{figure*}

In practice, excellent results are obtained after only a single iteration.
Indeed, since the CCD filling and readout models are unlikely to match the real
hardware better than $\mathcal{O}(\delta^3)$, only the first iteration is even
useful. To speed the correction of the 2368 COSMOS exposures, we therefore restricted
our correction of the larger data set to a single iteration. Figure~\ref{fig:correctedacsimage} shows the image from
figure~\ref{fig:rawacsimage} after correction, using one iteration.

\subsection{Additional features of the software} \label{sec:extracode}

Our CCD readout software has several additional features that we did not find necessary for CTI
correction in ACS/WFC, but which may be useful for other instruments. We document these features here for
completeness and future reference.

Although we used only two species of charge trap, the code is capable of handling many (currently, up to
four) different species, each with a density $\rho_q$ and characteristic release time $\tau$. With
additional measurements of flux loss due to long-lifetime charge traps, this could be used to more accurately
correct photometric measurements of astronomical sources.

The code can shuffle electrons both in the parallel direction and along a serial register.  In this mode,
the execution time is doubled.  Charge traps in the serial register, with characteristic release times
similar to the serial clock speed, can create trails at $90^\circ$ to those described by
\ref{eqn:trail}. However, it is well established \citep[\eg][]{riess03,mutchler05} that the serial CTE in the
ACS/WFC CCDs is very high, and that trailing is almost exclusively in the parallel direction. Indeed, we
attempted to measure serial CTI trails but found a signal consistent with zero. We therefore neglected
this entire mechanism.

By default, the code places model charge traps randomly within the model CCD. These are not at the same
locations as the charge traps in the real CCD, which introduces additional noise during the correction.
This is ameliorated by increasing the density of traps but decreasing their capacity as described in \S\ref{sec:forward}.
However, radiation-hardened detectors in future cameras \citep[\eg][]{dawson08} may contain sufficiently
few traps for this noise to become significant. If their individual locations could be measured on orbit, for example via pocket pumping, the
code can place model traps at the correct locations.

\begin{figure*}
\includegraphics[angle=90,width=128mm]{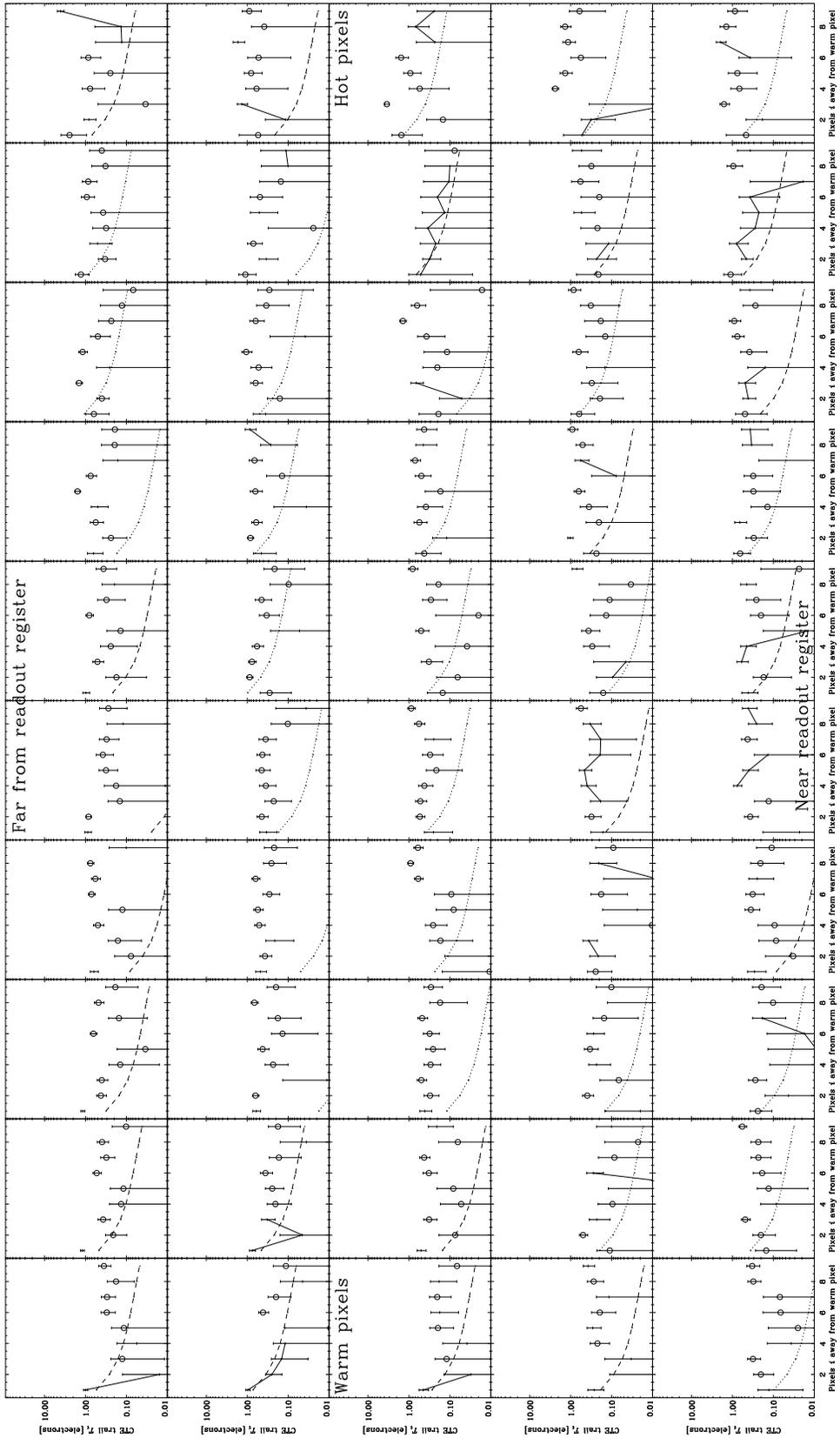}
\caption{Measurement of the CTI trails from the same images used in
figure~\ref{fig:rawtrails}, after correction. 
The $y$ axes are logarithmic; open circles and dotted (rather than dashed) lines show negative values.
}
\label{fig:correctedtrails}
\end{figure*}

\section{TESTING THE CORRECTION}\label{sec:tests}

\subsection{Warm pixel trails}

Figure~\ref{fig:correctedtrails} shows the residual CTI trails measured from corrected versions of
the images used for figure~\ref{fig:rawtrails}. Note the reduction in the level of all trails by
factor of about 30. This reduction is matched or bettered during all four of the times (the first,
sixth, seventh and thirteenth data points in figure~\ref{fig:trapgrowth}) when many COSMOS
exposures happened to be obtained during a single day, and a more precise measurement could be made. 
This demonstrates that we have achieved an approximately $(1-1/30)=97\%$ 
precision in our trapping and readout model, at all locations on the CCD array and for traps at
all heights within the CCD.

If the images are instead corrected using two iterations of the method in
table~\ref{tab:bristowmethod}, the data points in figure~\ref{fig:correctedtrails} move only
within their error bars, justifying our use of only one iteration for the sake of speed.

\subsection{Galaxy photometry measurement}

We now stack the corrected images to create 577 ``{\tt \_sci}'' science frames, following the pipeline
of \cite{koekemoer07}, which
makes use of the Drizzle/Multidrizzle software \citep{drizzle1,drizzle2}. There are four exposures per pointing, dithered almost along a straight
line (two exposures separated by $6.1\arcsec$ at an angle of $91.2^\circ$ from the coordinate
frame and another pair, separated by $3.1\arcsec$ from the first, at an angle of $85.4^\circ$).
These are transformed into a projection of the pixel array onto the sky, with oversampled
$0.03\arcsec$ pixels that are aligned with the instrument coordinate frame to keep the two readout
registers at the top and bottom. The dithers complicate the issue of how far each galaxy is
from the readout register, because it was in different areas of the focal plane in different
exposures. To roughly divide the catalogue into objects at final positions $(x,y)$ that have
traversed different numbers of charge traps, we calculate
\begin{equation} n_\mathrm{transfers}=\mathrm{max}\left\{0, 2048-|6046-1.67y-0.0614x|\right\} 
\end{equation} \noindent and separately consider objects with values of $n_\mathrm{transfers}$
in the ranges 0--475, 475--950, 950--1425, 1435--1900 and 1900--2048. The final bin also includes most of the area of the
oversampled images in which only data was available from only three of the four exposures, due to
the gap between the CCDs.

\begin{figure}
\includegraphics[width=84mm]{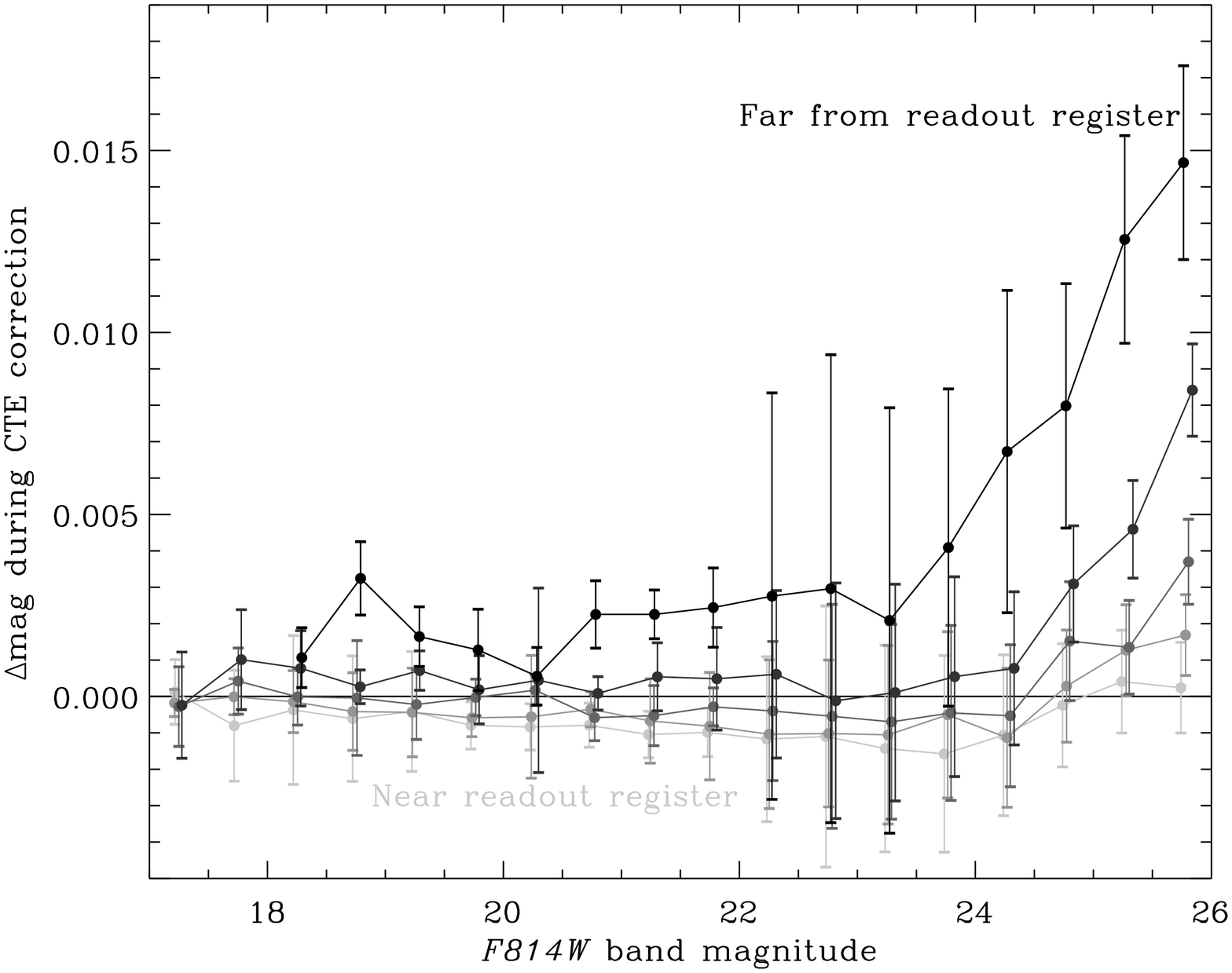}
\includegraphics[width=84mm]{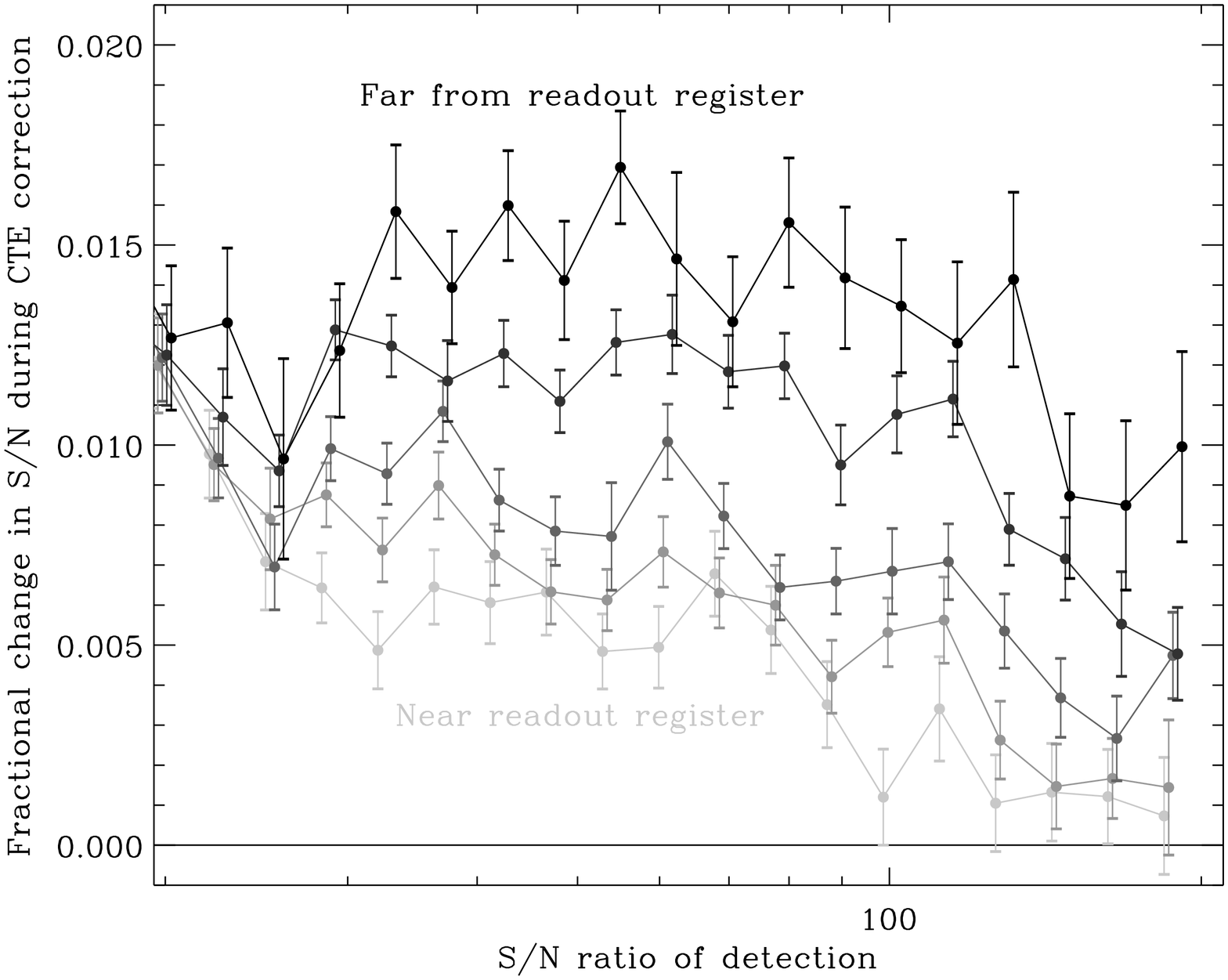}
\caption{Changes in galaxy photometry during CTI correction, as a function of the measured magnitude and detection S/N in the
corrected images. A fractional change is calculated as (old$-$new)/new. The five lines connect subsets of galaxies between approximately 
0--475, 475--950, 950--1425, 1435--1900 and 1900--2048
transfers from the readout register; the photometry is measured on stacked exposures and the exact number of transfers varies 
slightly between dithers. The points show the median value in the top panel, since the distribution of points is highly skewed, and the
mean value in the bottom panel, which is less so. Error bars show $68\%$ confidence limits on the mean.}
\label{fig:corrected_photometry}
\end{figure}

Figure~\ref{fig:corrected_photometry} shows the changes to galaxy photometry during CTE correction, as measured on original versus
corrected images by {\it Source Extractor} \citep{sex}. As expected from \cite{riess03}, we find that the most affected sources are
faint galaxies far from the readout register.  The amount of change is in line with the extrapolated prediction of \cite{riess04}.
Note however that \emph{this should not be interpreted as the error in point-source photometry on uncorrected images}. In particular,
our correction scheme does not currently include any species of charge traps with very long release times, which may steal additional
flux from sources. Our resolved sources also contain additional electrons in extended wings that, according to our well filling model,
lie near the bottom of the pixel potential, occupy a relatively large volume of silicon and are exposed to a greater number of charge
traps.

\begin{figure}
\includegraphics[width=84mm]{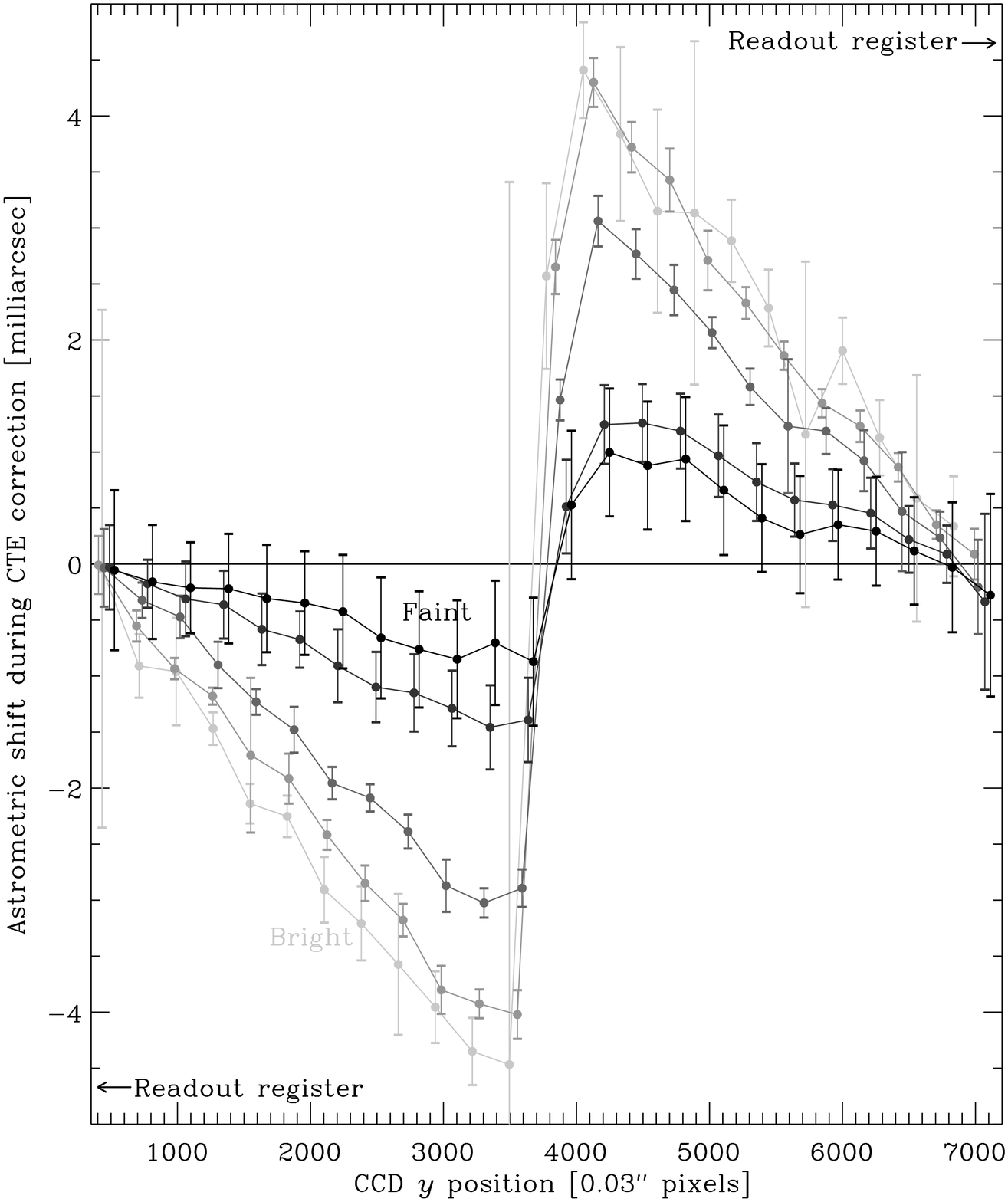}
\caption{Change in galaxy astrometry during CTI correction, in the direction of the readout registers. Readout registers are located
at opposite edges of the array, with pixels from each half being read out at their nearest register. Solid lines connect subsamples
of galaxies between magnitude limits of $F814W$~=~17--19, 19--21, 21--23, 23--25 and 25--27. For the sake of clarity in this
figure, the points show medians in each bin. The error bars show $68\%$ confidence limits, and the mean values behave consistently
around these bounds, but noise renders the overall behaviour less clear.
Note that faint galaxies are also typically smaller than bright galaxies, unlike the point sources used in previous studies.}
\label{fig:corrected_astrometry}
\end{figure}

Photometric measurements can also be expressed in terms of the galaxies' detection signal to noise. The interpretation for galaxies
with S/N$<20$ is complicated by two issues. Firstly, a selection bias in which sources whose detection signal to noise increased
greatly were not included in the pre-correction catalogue. Secondly, the correction is a form of unsharp masking, and noise in
adjacent pixels becomes less correlated during correction. As discussed in \cite{cti2}, this can actually lower the detection S/N of
faint sources.

\subsection{Galaxy astrometry measurement}

The change in galaxies' positions during CTI correction is shown in figure~\ref{fig:corrected_astrometry}. 
The spurious shift is $2.42\pm 0.17\times 10^{-6}\arcsec$ per ($0.05\arcsec$) pixel transfer for galaxies between
magnitudes 17 and 19, and $0.57\pm 0.24\arcsec$ per transfer for $25^\mathrm{th}$ to $27^\mathrm{th}$ magnitude
galaxies.

The distance depends as expected upon distance from the readout register, but is initially surprising as a function of galaxy flux.
Using very similar code, \cite{bristow04} found bright point sources to be less affected than faint ones. Our own code also moves a
given object less if its flux is increased. However, bright resolved sources also tend to be larger than faint ones. Astrometry is
most affected by charge traps with short release times \citep{cti2}, which are now able to capture, move and release an electron
several times within a large galaxy. Furthermore, the non-linear well filling model ensures that even electrons in the low-level,
extended wings are exposed to most of the charge traps seen by the core. Multiple trapping does not remove flux from a galaxy, so
should not impact photometric measurements. However, this process breaks the standard argument whereby charge traps saturate, with
empty traps being newly exposed only by relatively large increases in flux. Subtle efects like this highlight the power of a
pixel-based method: they are unlikely to be matched in a parametric scheme acting at a catalogue level.

\subsection{Galaxy shape measurement}

\begin{figure}
\includegraphics[width=84mm]{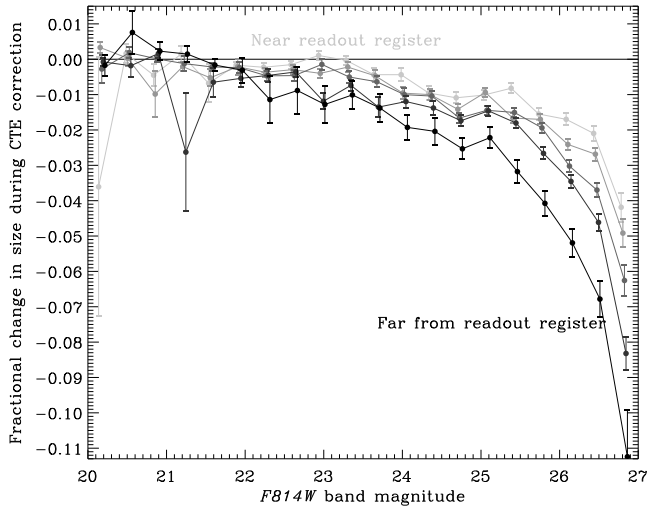}
\caption{Change in galaxy FWHM size during CTI correction. Solid lines connect the same samples as in 
figure~\ref{fig:corrected_photometry}. Points show mean values within bins and $68\%$ confidence limits.}
\label{fig:corrected_size}
\end{figure}

Figure~\ref{fig:corrected_size} shows the change in galaxies' full width at half maximum size during CTI correction, as measured
by {\it Source Extractor} \citep{sex}. The CTI trails had spuriously enlarged objects, which are shrunk slightly when the flux in
the trails is pushed back into the core. The effect increases as expected with the number of charge transfers, and is most
pronounced in faint galaxies, which are intrinsically smallest, and also encounter the largest number of charge traps per
electron during readout.

Galaxy ellipticities are a particularly interesting case because
the {\it true} value of ellipticity is known, at least statistically. If there is no preferred
direction in the Universe, galaxies cannot preferentially align in any direction and their mean ellipticity ought to be zero.
However, the CTI trails coherently elongate galaxies in the readout direction. 

\begin{figure}
\includegraphics[width=84mm]{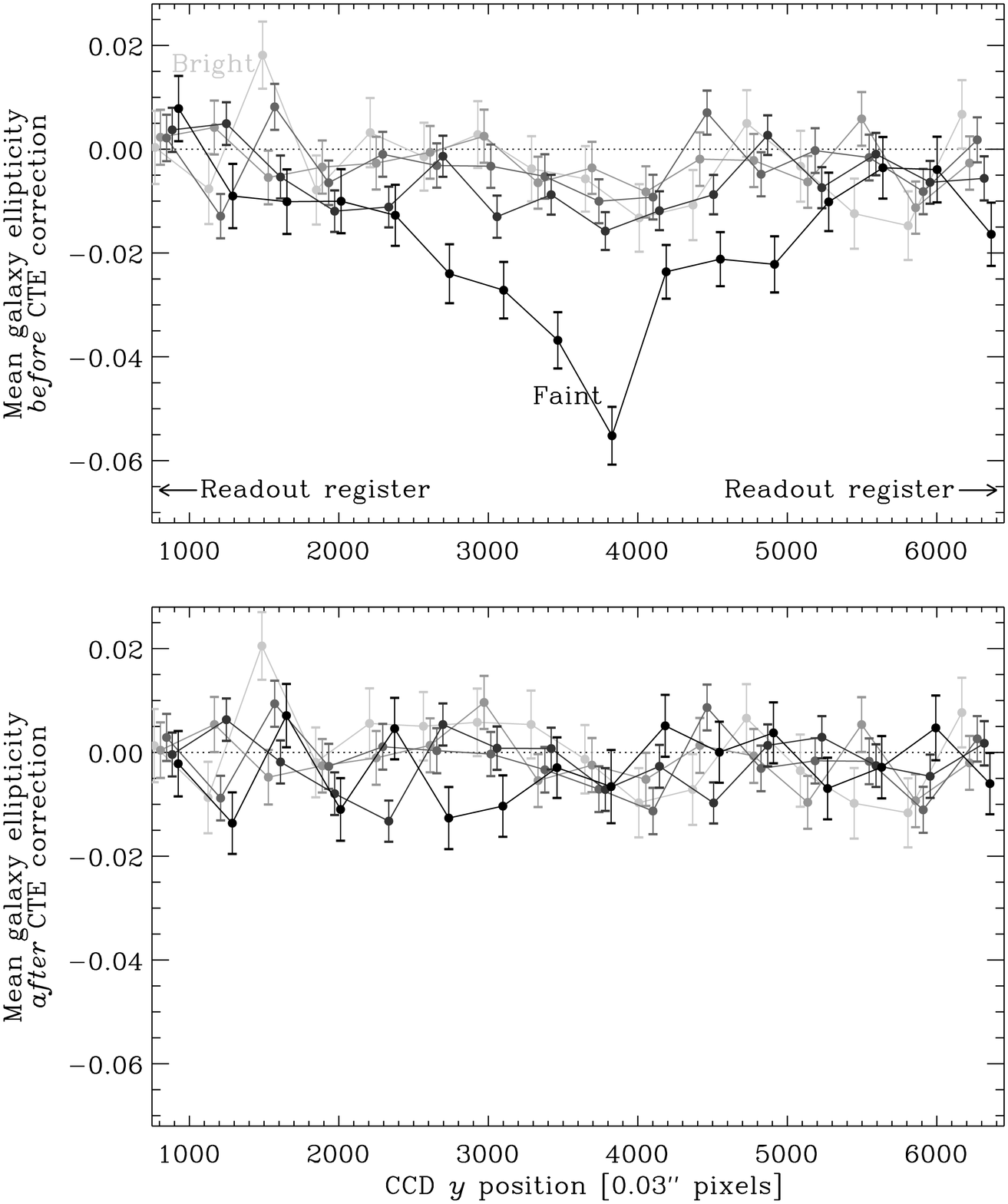}
\caption{Component of mean galaxy ellipticities in the direction of the readout registers ($e\cos{2\theta}$, where $e$ is the
ellipticity and $\theta$ is the angle between the major axis and the line joining opposite sides of the CCD) before and after CTI
correction. Values are shown after correction for convolution with the PSF, whose shape is imprinted on the galaxies', but this
correction increases the scatter. Solid lines connect subsamples
of galaxies within magnitude limits of $F814W$~=~22--23, 23--24, 24--25, 25--26 and 26--27. 
Points show mean values in bins and $68\%$ confidence limits.}
\label{fig:corrected_ellipticity}
\end{figure}

In practice, high precision measurements of galaxy shapes are made difficult because they are imprinted with the shape of the
telescope's PSF. Furthermore, the ACS/WFC PSF changes during orbit, as HST expands or contracts due to thermal shifts. Most
inconveniently, the predominant PSF patterns happen to align themselves with the readout direction and also tend to align
galaxies in that direction \citep{rhodes07}. Fortunately, a great deal of software has been developed for effectively
``deconvolving'' galaxy ellipticities from the PSF, in order to measure weak gravitational lensing. We use the algorithm by
\cite{rrg}, as implemented by \cite{leauthaud07} and \cite{massey07cs}, to recover the underlying galaxy ellipticities. 

The top panel of figure~\ref{fig:corrected_ellipticity} shows how galaxies appear spuriously elongated in the direction of the
readout registers before CTE correction. The faintest galaxies are worst affected, although the elongation is also seen in the
brightest galaxies. As shown in \cite{cti2}, the effect is also worst for galaxies intrinsically elongated perpenicular to the
readout direction. Neither such effect is modelled by the parametric correction of \cite{rhodes07}. The mean spurious
ellipticity of galaxies in the faintest two bins increases as approximately $(24.3\pm 2.9)\times10^{-6}$ per ($0.05\arcsec$
pixel) transfer and $(7.7\pm 2.0)\times10^{-6}$ per transfer. The noise is predominantly due to the process of PSF
deconvolution. After CTI correction, as shown in the bottom panel, this effect has been successfully reduced to $(0.2\pm
3.1)\times10^{-6}$ and  $(2.2\pm 2.1)\times10^{-6}$. The mean shear is similarly reduced by an order of magnitude from
$(1.5\pm0.1)\%$ to $(0.2\pm0.2)\%$ in the faintest bin, and from $(7.4\pm1.1)\times10^{-6}$ to $(0.9\pm1.1)\times10^{-6}$
overall.

\section{DISCUSSION \& CONCLUSIONS}\label{sec:conclusions}

We have corrected Charge Transfer Inefficiency trailing in ACS/WFC images by more than an order of magnitude. As demonstrated
by the improvement of HST COSMOS survey images between figures~\ref{fig:rawtrails} and \ref{fig:correctedtrails}, and in
figure~\ref{fig:corrected_ellipticity},  we have achieved a 90-99\% precision in our model, at all locations on the CCD array
and for traps at all heights within the CCD. Our method works at the pixel level, moving individual electrons back to locations
from which they were dragged during readout, and model parameters correspond to real, physical properties of the device. This
is far better than the ad-hoc, parametric correction described by \citet{rhodes07}. Since the trails are created during the
final step of data acquisition, the correction should be applied as the first step in data reduction and analysis.


As suggested by \citet{biretta05}, warm pixels in the detectors were successfully used to measure
device properties including the charge trap density and release times. The ubiquitous
distribution of warm pixels allows the locations of charge traps to be continuously measured across the CCD,
using in-orbit science data. Exceeding earlier expectations, however, the range of flux levels in warm
pixels can also be used to obtain the 3D height of the traps, and the rate at which electrons fill
up the volume of a pixel's potential well. All of these parameters are needed to correct CTI
trailing, and it is remarkable that they can all be measured using a separate consequence of the
radiation damage. To improve on this with future large surveys will probably require laboratory-based
measurements of the CCD and charge trap characteristics before launch (and possibly after
retrieval), or the integration of on-board electronics to allow in-flight tests such as pocket
pumping.

We find two significant species of charge traps that affect the astrometry and morphology of
astronomical objects, with characteristic release times $\tau=10.4\pm3.2$~pixels and
$\tau=0.88\pm0.2$~pixels. These accumulated in a ratio of approximately 3.0:1 and, by the end of
HST cycle 13, there was slightly more than one in every other pixel, uniformly distributed
throughout the CCD. An electron beginning 2048 pixels away from the readout register therefore
encounters a large number of traps during readout. The (random) locations of traps in our model
are inevitably different to those in the real CCD. This adds shot noise during correction that is
particularly noticeable in the structure of the image background. We have eliminated half of the
shot noise by increasing the density of model traps density and allowing them to hold non-integer
amounts of charge.

We did not attempt to measure the density of traps with very long release times. As shown by
\cite{cti2}, these primarily affect object photometry. They can be measured using repeated images
of standard stellar fields, such as the STScI internal CTE calibration programme. Alternatively,
once the uniform distribution of other charge trap species has been established, they could also
be measured from EPER or FPR tests. If their density is measured after servicing mission 4,
our software can easily handle additional species of charge traps, and thus improve the simultaneous
correction of object photometry, astrometry and morphology.

We measure a functional form describing the rate at which electrons fill up the volume of a
pixel's potential well. Below a ``notch'' depth of $96.5\pm2.0$ e$^-$, they occupy negligible
volume; additional electrons overflow and the cloud expands with a power law of index
$\alpha=0.576\pm0.013$. This volume determines the number of traps encountered by the cloud, as it
is moved through the CCD substrate towards the readout register. Note that, in order to cope with
the huge data volume, we developed the model stepwise: fixing the first measured parameters before
fitting the later ones. Since errors on some measurements were not carried through, these later
errors are likely to be underestimated.

Our pixel filling model neatly illustrates why CTI trailing is a non-linear function of object
flux, and why it can be reduced by pre- or post-flashing an image to increase the background
level. During readout, the large cloud of electrons within a bright source has to navigate more
charge traps that the small cloud within a faint source -- but the ratio of exposed traps to
electrons falls as $n_q/n_e\approx n_e^{\alpha-1}$. Since $\alpha<1$, the fractional amount of
charge that gets trailed behind a faint source is greater than that behind a bright source. In
addition, a fat zero reduces the {\em total} amount of trailing behind all sources by moving all
pixels to the right in figure~\ref{fig:qn}. For faint sources of fixed flux, the total number of
exposed traps is proportional to ${\rm d}n_q/{\rm d}n_e$, which decreases as
$\sim(b-d)^{\alpha-1}$.

The two biggest simplifications in our model are
reducing the three-phase clocking cycle to a single shift, and not explicitly treating the
continuous injection of charge from warm pixels during readout. The latter would be particularly
important if this method were required for measurements of charge traps with long release times. 
Improvements in these areas would be interesting directions for future studies.

\section*{Acknowledgments}

The authors gratefully thank Chris Bebek, Paul Bristow, Marissa Cevallos, Nick Cross, Kyle Dawson, Andy Fruchter,
Nigel Hambly, Mike Lampton, Michael Levi, Max Mutchler, Natalie Roe, George Seabroke, 
Suresh Seshadri, Patrick Shopbell, Tim Schrabback and Roger Smith for sharing their expertise.
The HST ACS CTI calibration program is funded through NASA grant HST-AR-10964.
The HST COSMOS Treasury program was funded through NASA grant HST-GO-09822 with P.I.\ Nick Scoville. 
RM is supported by STFC Advanced Fellowship \#PP/E006450/1 and FP7 grant MIRG-CT-208994.
AL acknowledges support from the Chamberlain Fellowship at LBNL and from the Berkeley Center for Cosmological Physics.

This work was based on observations with the NASA/ESA {\em Hubble Space Telescope}, obtained at the Space Telescope
Science Institute, which is operated by AURA Inc, under NASA contract NAS 5-26555.
W

\appendix

\section{CCD anneal dates}

\begin{table}
\begin{center}
\vspace{3mm}
\begin{tabular}{r@{~}l@{\,}lr@{:}c@{:}l}
\hline
\hline
  9&Sep&2003&08&34&58\\
 12&Oct&2003&07&09&36\\
  6&Nov&2003&15&11&06\\
  1&Dec&2003&15&48&13\\
  4&Jan&2004&09&01&30\\
 30&Jan&2004&01&08&26\\
  4&Mar&2004&01&21&04\\
 27&Mar&2004&00&59&38\\
 25&Apr&2004&21&31&00\\
 22&May&2004&07&34&57\\
 18&Jun&2004&04&36&37\\
  6&Jul&2004&16&28&51\\
 14&Jul&2004&16&52&05\\
 10&Aug&2004&04&41&59\\
  8&Sep&2004&06&01&41\\
  8&Oct&2004&09&05&45\\
  5&Nov&2004&19&42&53\\
  2&Dec&2004&14&11&50\\
 30&Dec&2004&12&40&00\\
 29&Jan&2005&11&00&00\\
 12&Feb&2005&09&04&52\\
  4&Mar&2005&22&55&30\\
 24&Mar&2005&09&12&24\\
 19&Apr&2005&07&33&57\\
 20&May&2005&05&00&00\\
 12&Jun&2005&22&35&22\\
 16&Jul&2005&07&58&20\\
 11&Aug&2005&20&37&49\\
  9&Sep&2005&08&45&56\\
  8&Oct&2005&10&06&47\\
  3&Nov&2005&15&20&00\\
 25&Nov&2005&08&37&56\\
 31&Dec&2005&04&30&00\\
\hline
\hline
\end{tabular}
\caption{Dates and times when the ACS detectors were annealed during HST cycles 12 and 13.
\label{tab:acsanneals}}
\end{center}
\end{table}

\label{lastpage}

\end{document}